\documentstyle[12pt,axodraw]{article}
\begin{document}
\newcommand{\bq}{\begin{equation}}
\newcommand{\eq}{\end{equation}}
\newcommand{\bqa}{\begin{eqnarray}}
\newcommand{\eqa}{\end{eqnarray}}
\newcommand{\nl}{\nonumber \\}
\newcommand{\epl}{e^+}
\newcommand{\emn}{e^-}
\newcommand{\ubar}{\bar{u}}
\newcommand{\dbar}{\bar{d}}
\newcommand{\cbar}{\bar{c}}
\newcommand{\sbar}{\bar{s}}
\newcommand{\bbar}{\bar{b}}
\newcommand{\si}{\sigma}
\newcommand{\sit}{\sigma_{tot}}
\newcommand{\sqs}{\sqrt{s}}
\newcommand{\p}[1]{{\scriptstyle{\,(#1)}}}
\newcommand{\gev}{\mbox{GeV}}
\newcommand{\mev}{\mbox{MeV}}
\title{
\vspace{-4cm}
\begin{flushright}
{\large  INLO-PUB-8/94}
\end{flushright}
\vspace{4cm}
Four quark processes at LEP 200
\footnote{This research has been partly supported by EU under contract
number\\ CHRX-CT-0004.}}
\author{R.~Pittau\thanks{email address: pittau@rulgm0.LeidenUniv.nl}\\
        Instituut-Lorentz, Leiden, The Netherlands}
\maketitle
\begin{abstract}
In this paper I investigate the production of four quarks at LEP 200 energies.
Effects due to initial state QED corrections and background
diagrams, including QCD contributions, are studied and examples of results
obtained with an event generator presented.
\end{abstract}
\clearpage
\setcounter{page}{1}
\section{Introduction}
At LEP 200, four jet events will be used to perform
a direct reconstruction of the $W$ massl.gov    \cite{katsa}.
The mass of the charged boson can be determined by looking at the maximum
of the detected invariant mass distribution in the process
\bq
\epl\;\emn \to W^+\;W^- \to 4~quarks \label{wpair}\;\;.
\eq
Since the energy of the jets is not accurately known, one has to perform
fits to the measured distribution using, as a constraint, information about
the center of mass energy of the colliding electrons. This knowledge is
crucial.
If $\epsilon$ is the average energy radiated by the beams, one expects
a shift
\bq
\Delta_M=\epsilon\,{M_W \over \sqs}\;\;.
\eq
in the reconstructed mass.
The value of the maximum of the theoretically predicted distribution is
not affected by
initial state QED radiation or by the inclusion of electroweak and QCD
background diagrams, but the average energy loss is very sensitive
to them and to the experimentally applied cuts.
Therefore, taking into account only the three basic diagrams of reaction
(\ref{wpair}) is not sufficient for comparison with experiment and a flexible
tool is needed, that includes
\begin{itemize}
\item the possibility to implement any desired cut and to study any
experimental distribution;
\item all electroweak background diagrams, leading to the same four quark
final state;
\item initial state QED radiation;
\item QCD background diagrams.
\end{itemize}
In ref. \cite{cpc1} a Monte Carlo program was described that satisfies
all these requirements. The inclusion of electroweak background diagrams
for {\em all} possible four fermion final states was performed in ref.
\cite{np}, and QED initial state radiation introduced in ref. \cite{qed}.
Here, I discuss the effects of QCD production channels on four
quark processes, also showing numerical results obtained with
the above Monte Carlo.
\section{Implementing QCD contributions}
There are two types of QCD contributions. On one hand, when considering a
specific four quark final state, the particles can be produced either
by decaying gauge bosons or gluons, and the respective terms interfere
when squaring the amplitude. Therefore, this gluonic contribution
constitutes an {\em interfering} QCD background for the process (\ref{wpair}).
On the other hand, the measured signal is a four jet final
state, thus also the {\em non-interfering} cross section due to
$\epl \emn \to g g q \bar{q}$ has to be included.

The implementation of the interfering QCD diagrams is trivial
in a program like that in ref. \cite{cpc1}, which already contains
{\em all} possible electroweak Feynman diagrams.
It is enough to add gluons wherever photons connect quark lines.
In fact, if $Q_q,~Q_{q'}$ are the charges of the quarks and $i,j,l,m$
colour labels, the amplitude due to photon
exchange between quarks can be written as follows
\bq
{\cal M}_\gamma= \alpha^2 \, Q_q\, Q_{q'} \left[\,{\cal A}^\p{1}\delta_{ij}
\delta_{lm}+{\cal A}^\p{2}\delta_{im}\delta_{jl}\,\right]\;\;.
\eq
Substituting
\bqa
\delta_{ij}\delta_{lm} &\to& \left(1-{\alpha_r \over 6}\right)\,
                             \delta_{ij}\delta_{lm} +
                              {\alpha_r \over 2}\,\delta_{im}\delta_{jl} \nl
\delta_{im}\delta_{jl} &\to& \left(1-{\alpha_r \over 6}\right)\,
                             \delta_{im}\delta_{jl} +
                              {\alpha_r \over 2} \,\delta_{ij}\delta_{lm} \nl
\alpha_r &=& { \alpha_s \over \alpha\, Q_q\, Q_{q'}} \nonumber
\eqa
in the previous formula takes also gluons into account, including the correct
QCD coupling and colour structure.
Since there is no need to evaluate new Feynman diagrams, this approach requires
almost no additional CPU time.

As for the the non-interfering $\epl \emn \to g g q \bar{q}$ process, the
relevant amplitude can be efficiently computed by using the recursion relations
of ref. \cite{gilber}. A Monte Carlo program with the same structure of that of
ref. \cite{cpc1} has been built in order to get this contribution.

Finally, comparisons have been made with results of refs. \cite{maina,mainap},
always finding a very good agreement.
\section{Results}
I shall now discuss a number of results obtained with the program
of ref. \cite{cpc1}. Fragmentation effects are neglected.
The chosen values for the input parameters are
$\alpha= 1/129$, $\alpha_s$= 0.103, $\sin^2\theta_W= 0.23$, $M_W= 80.5$,
$\Gamma_W= 2.3$, $M_Z= 91.19$ and $\Gamma_Z= 2.5$ (all GeV).

\begin{table}[h]
\begin{center}
\begin{tabular}{|l|cc|c|} \hline\hline
$\sqrt s$ (GeV) & \multicolumn{2}{c|}{175}  &  190   \\
\hline
    & $\sigma$ &$\epsilon$& $\sigma$  \\
\hline
$WW$ signal                        &3.0674 &--     &3.5136 \\
                                   &0.0074 &       &0.0090 \\
\hline
$WW$ signal + ISR                  &2.5622 &1.091  &3.1416 \\
                                   &0.0071 &0.005  &0.0089 \\
\hline
 All EW diagrams + ISR             &2.5922  &1.104 &3.3553 \\
                                   &0.0075  &0.005 &0.0097 \\
\hline
 All EW diagrams + interfering QCD & 2.6202 &1.188 &3.3789 \\
   + ISR                           &0.0079  &0.017 &0.0100 \\
\hline
 All EW diagrams + all QCD         &3.1155  &2.513 &3.8688 \\
    + ISR                          &0.0123  &0.065 &0.0146 \\
\hline \hline
\end{tabular}
\end{center}
\caption[.]{Cross section in picobarn and average energy loss ($\epsilon$)
in GeV for $e^+ e^- \to$ 4 $jets$. $E_\p{all~particles}>20~GeV$,
$|\cos \theta_{\p{all~particles}}|<0.9$, $m\p{ij}>10~GeV$ and
$|\cos \p{i,\,j}|<0.9$ between all possible final state pairs.
The second line in each entry is the estimated Monte Carlo error.}
\label{tableone}
\end{table}

Table \ref{tableone} shows results for $e^+ e^- \to 4~jets$
under inclusion of several QED and QCD contributions at LEP 200 energies. The
cross section is lowered by the initial state QED radiation (ISR),
while all the other contributions tend to raise it back to its Born value.
The electroweak (EW) background is at the per cent level as well as the
interfering QCD background. In the last row
also $e^+ e^- \to 2\,gluons\,+\,2\,quarks$ is included; it increases the cross
section by 16\% at $\sqrt s$= 175 GeV.
\noindent As far as the energy loss is concerned,
QCD interfering diagrams are more important than
EW background. However, both contributions are
small compared to the huge effect due to the inclusion of $ggq\bar{q}$.
Cuts on the invariant masses around the $W$ mass strongly suppress these QCD
events. For instance, with the additional requirement that at least one
invariant mass falls in the interval $M_W \,\pm\, 2.5$ GeV, the $ggq\bar{q}$
cross section goes down by a factor 4 and the $WW$ signal only by 5\%.
Furthermore, the corrections to the energy loss become of the same order of
magnitude as the interfering QCD contributions. Finally, it should be noted
that $\epsilon$ is also very sensitive to the experimental setup \cite{qed}.
For example, with the additional cut described above, the value of $\epsilon$
in the second row of table \ref{tableone} goes down from
1.091 $\pm$ 0.005 to 1.039 $\pm$ 0.007.

In figure 1 the effect of the combinatorial background on the invariant mass
distribution for a specific final state at $\sqs=$ 175 GeV is shown. In (a) any
possible invariant mass $m\p{i,j}$ is taken into account, so that the
normalization of the histogram is 6 times the value of the cross section.
In (b) only
those combinations that can be produced by decaying $W's$ are plotted.

Figure 2 shows invariant mass distributions in four jet events under
inclusion of QCD, QED and EW background contributions at two different
energies.
The curves have been obtained by fitting histograms as those given in figure 1.
ISR lowers the distributions and $g g q \bar{q}$ gives a roughly constant
positive contribution between 30 and 100 GeV. Below the peak the latter effect
is dominant, while at the peak the situation is reversed.

A more detailed analysis of the peak of the distribution at $\sqs=$ 175 GeV
is contained in figure 3. The maximum is not affected by ISR, EW background and
QCD contributions, and, after subtraction of constant effects due to
combinatorial background and  $g g q \bar{q}$ production, also the width
is unchanged.
\section{Conclusions}
At LEP 200 the $W$ mass will be measured from invariant mass distributions
by direct reconstruction.
An accurate analysis shows that, in four jet events, the maximum and
the width of the distributions are not strongly affected  by the inclusion of
electroweak and QCD background diagrams as well as QED initial state radiation.
On the other hand, due to uncertainties in the measurement of the jet energy,
a precise knowledge of the average energy loss is required, and this quantity
{\em is very sensitive} to the mentioned effects and to the experimental setup.
Therefore, a flexible program like that in ref. \cite{cpc1} is needed, which
includes all these corrections and the possibility to implement cuts and
compute any experimental distribution.
\section{Acknowledgments}
Very stimulating discussions with Prof. Berends and Dr. Kleiss are gratefully
acknowledged. I also thank Dr. Ballestrero, Dr. Maina and Dr. Moretti for
detailed information on the results of ref. \cite{maina} and for checks against
their programs.

\begin{picture}(180,365)(40,-45)
\LinAxis(0,0)(200,0)(2,10,5,0,1)
\LinAxis(0,300)(200,300)(2,10,-5,0,1)
\LinAxis(0,0)(0,300)(6,10,-5,0,1)
\LinAxis(200,0)(200,300)(6,10,5,0,1)
\SetScale{100.}
\SetWidth{0.005}
\Line(0.0000,0.0000)(0.1000,0.0000)\Line(0.1000,0.0000)(0.1000,0.0998)
\Line(0.1000,0.0998)(0.2000,0.0998)\Line(0.2000,0.0998)(0.2000,0.4435)
\Line(0.2000,0.4435)(0.3000,0.4435)\Line(0.3000,0.4435)(0.3000,0.5818)
\Line(0.3000,0.5818)(0.4000,0.5818)\Line(0.4000,0.5818)(0.4000,0.7067)
\Line(0.4000,0.7067)(0.5000,0.7067)\Line(0.5000,0.7067)(0.5000,0.7596)
\Line(0.5000,0.7596)(0.6000,0.7596)\Line(0.6000,0.7596)(0.6000,0.7382)
\Line(0.6000,0.7382)(0.7000,0.7382)\Line(0.7000,0.7382)(0.7000,1.5667)
\Line(0.7000,1.5667)(0.8000,1.5667)\Line(0.8000,1.5667)(0.8000,2.0720)
\Line(0.8000,2.0720)(0.9000,2.0720)\Line(0.9000,2.0720)(0.9000,0.4174)
\Line(0.9000,0.4174)(1.0000,0.4174)\Line(1.0000,0.4174)(1.0000,0.2434)
\Line(1.0000,0.2434)(1.1000,0.2434)\Line(1.1000,0.2434)(1.1000,0.0857)
\Line(1.1000,0.0857)(1.2000,0.0857)\Line(1.2000,0.0857)(1.2000,0.0077)
\Line(1.2000,0.0077)(1.3000,0.0077)\Line(1.3000,0.0077)(1.3000,0.0000)
\Line(1.3000,0.0000)(1.4000,0.0000)\Line(1.4000,0.0000)(1.4000,0.0000)
\Line(1.4000,0.0000)(1.5000,0.0000)\Line(1.5000,0.0000)(1.5000,0.0000)
\Line(1.5000,0.0000)(1.6000,0.0000)\Line(1.6000,0.0000)(1.6000,0.0000)
\Line(1.6000,0.0000)(1.7000,0.0000)\Line(1.7000,0.0000)(1.7000,0.0000)
\Line(1.7000,0.0000)(1.8000,0.0000)\Line(1.8000,0.0000)(1.8000,0.0000)
\Line(1.8000,0.0000)(1.9000,0.0000)\Line(1.9000,0.0000)(1.9000,0.0000)
\SetScale{1.}\SetWidth{0.5}
\Text(10,250)[l]{${\mbox{pb} \over 10\,\gev}$}
\Text(20,280)[]{(a)}
\Text(0,-10)[]{0}\Text(100,-10)[]{100}
\Text(200,-10)[r]{200}
\Text(-14,50)[]{0.25}
\Text(-14,100)[]{0.50}
\Text(-14,150)[]{0.75}
\Text(-14,200)[]{1.00}
\Text(-14,250)[]{1.25}
\Text(100,-25)[t]{$m\p{i,j}$ (GeV)}
\end{picture}
\begin{picture}(180,365)(-10,-45)
\LinAxis(0,0)(200,0)(2,10,5,0,1)
\LinAxis(0,300)(200,300)(2,10,-5,0,1)
\LinAxis(0,0)(0,300)(6,10,-5,0,1)
\LinAxis(200,0)(200,300)(6,10,5,0,1)
\SetScale{100.}
\SetWidth{0.005}
\Line(0.0000,0.0000)(0.1000,0.0000)\Line(0.1000,0.0000)(0.1000,0.0006)
\Line(0.1000,0.0006)(0.2000,0.0006)\Line(0.2000,0.0006)(0.2000,0.0015)
\Line(0.2000,0.0015)(0.3000,0.0015)\Line(0.3000,0.0015)(0.3000,0.0037)
\Line(0.3000,0.0037)(0.4000,0.0037)\Line(0.4000,0.0037)(0.4000,0.0057)
\Line(0.4000,0.0057)(0.5000,0.0057)\Line(0.5000,0.0057)(0.5000,0.0128)
\Line(0.5000,0.0128)(0.6000,0.0128)\Line(0.6000,0.0128)(0.6000,0.0436)
\Line(0.6000,0.0436)(0.7000,0.0436)\Line(0.7000,0.0436)(0.7000,0.9409)
\Line(0.7000,0.9409)(0.8000,0.9409)\Line(0.8000,0.9409)(0.8000,1.5436)
\Line(0.8000,1.5436)(0.9000,1.5436)\Line(0.9000,1.5436)(0.9000,0.0194)
\Line(0.9000,0.0194)(1.0000,0.0194)\Line(1.0000,0.0194)(1.0000,0.0019)
\Line(1.0000,0.0019)(1.1000,0.0019)\Line(1.1000,0.0019)(1.1000,0.0004)
\Line(1.1000,0.0004)(1.2000,0.0004)\Line(1.2000,0.0004)(1.2000,0.0002)
\Line(1.2000,0.0002)(1.3000,0.0002)\Line(1.3000,0.0002)(1.3000,0.0000)
\Line(1.3000,0.0000)(1.4000,0.0000)\Line(1.4000,0.0000)(1.4000,0.0000)
\Line(1.4000,0.0000)(1.5000,0.0000)\Line(1.5000,0.0000)(1.5000,0.0000)
\Line(1.5000,0.0000)(1.6000,0.0000)\Line(1.6000,0.0000)(1.6000,0.0000)
\Line(1.6000,0.0000)(1.7000,0.0000)\Line(1.7000,0.0000)(1.7000,0.0000)
\Line(1.7000,0.0000)(1.8000,0.0000)\Line(1.8000,0.0000)(1.8000,0.0000)
\Line(1.8000,0.0000)(1.9000,0.0000)\Line(1.9000,0.0000)(1.9000,0.0000)
\SetScale{1.}\SetWidth{0.5}
\Text(10,250)[l]{${\mbox{pb} \over 10\,\gev}$}
\Text(20,280)[]{(b)}
\Text(0,-10)[]{0}\Text(100,-10)[]{100}
\Text(200,-10)[r]{200}
\Text(-14,50)[]{0.25}
\Text(-14,100)[]{0.50}
\Text(-14,150)[]{0.75}
\Text(-14,200)[]{1.00}
\Text(-14,250)[]{1.25}
\Text(100,-25)[t]{$m\p{3,4}$ + $m\p{5,6}$ (GeV)}
\end{picture}

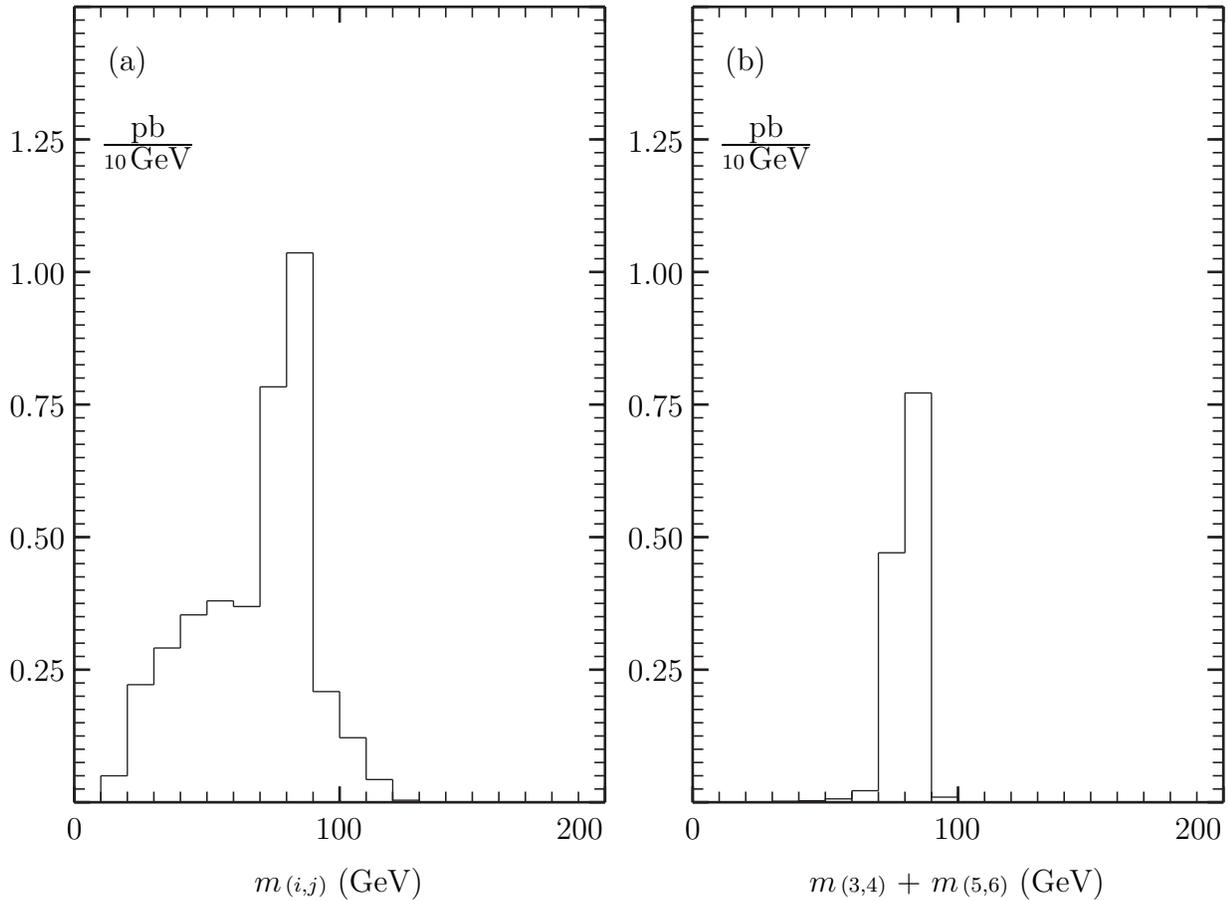
\noindent Figure 1: Distributions of the invariant masses for
$e^+ e^- \to$  $u\p{3}\,\bar{d}\p{4}\,d\p{5}\,\bar{u}\p{6}$
 with (a) and without (b) combinatorial background
at $\sqrt s$ = 175 GeV. Cuts like in table 1.
All diagrams are taken into account.
QED corrections and QCD background included.
\clearpage
\begin{picture}(180,365)(40,-45)
\LinAxis(0,0)(200,0)(2,10,5,0,1)
\LinAxis(0,300)(200,300)(2,10,-5,0,1)
\LinAxis(0,0)(0,300)(6,10,-5,0,1)
\LinAxis(200,0)(200,300)(6,10,5,0,1)
\SetScale{100.}
\SetWidth{0.005}
\DashCurve{(0.1,0.0000)
(0.2,0.1176)
(0.3,0.5182)
(0.4,0.6876)
(0.5,0.8400)
(0.6,0.8840)
(0.7,0.8480)
(0.8,1.8440)
(0.9,2.4980)
(1.0,0.5107)
(1.1,0.3207)
(1.2,0.1223)
(1.3,0.0111)
(1.4,0.0001)
(1.5,0.0000)
(1.6,0.0000)
(1.7,0.0000)
(1.8,0.0000)
(1.9,0.0000)
(2.0,0.0000)}{0.03}
\Curve{(0.1,0.0000)
(0.2,0.2079)
(0.3,0.6097)
(0.4,0.7529)
(0.5,0.8941)
(0.6,0.9365)
(0.7,0.9119)
(0.8,1.7490)
(0.9,2.2310)
(1.0,0.5408)
(1.1,0.3314)
(1.2,0.1454)
(1.3,0.0355)
(1.4,0.0015)
(1.5,0.0000)
(1.6,0.0000)
(1.7,0.0000)
(1.8,0.0000)
(1.9,0.0000)
(2.0,0.0000)}
\DashCurve{(0.1,0.0000)
(0.2,0.1021)
(0.3,0.1578)
(0.4,0.1586)
(0.5,0.1683)
(0.6,0.1641)
(0.7,0.1564)
(0.8,0.1631)
(0.9,0.1424)
(1.0,0.1073)
(1.1,0.0813)
(1.2,0.0564)
(1.3,0.0267)
(1.4,0.0015)
(1.5,0.0000)
(1.6,0.0000)
(1.7,0.0000)
(1.8,0.0000)
(1.9,0.0000)
(2.0,0.0000)}{0.013}
\SetScale{1.}\SetWidth{0.5}
\DashLine(105,250)(130,250){3}
\Text(10,250)[l]{${\mbox{pb} \over 10\,\gev}$}
\Text(20,280)[]{(a)}
\Text(135,250)[l]{$WW$}
\Line(105,235)(130,235)
\Text(135,235)[l]{All + QCD}
\Text(135,220)[l]{~~~~~+ ISR}
\DashLine(105,200)(130,200){1.3}
\Text(135,200)[l]{$g g q\bar{q}$}
\Text(0,-10)[]{0}\Text(100,-10)[]{100}
\Text(200,-10)[r]{200}
\Text(-10,50)[]{1}
\Text(-10,100)[]{2}
\Text(-10,150)[]{3}
\Text(-10,200)[]{4}
\Text(-10,250)[]{5}
\Text(100,-25)[t]{$m\p{i,j}$ (GeV)}
\end{picture}
\begin{picture}(180,365)(-10,-45)
\LinAxis(0,0)(200,0)(2,10,5,0,1)
\LinAxis(0,300)(200,300)(2,10,-5,0,1)
\LinAxis(0,0)(0,300)(6,10,-5,0,1)
\LinAxis(200,0)(200,300)(6,10,5,0,1)
\SetScale{100.}
\SetWidth{0.005}
\DashCurve{(0.10,0.0000)
(0.20,0.1308)
(0.30,0.5089)
(0.40,0.6987)
(0.50,0.8102)
(0.60,0.8320)
(0.70,0.8338)
(0.80,1.9387)
(0.90,2.8062)
(1.00,0.6351)
(1.10,0.4960)
(1.20,0.4036)
(1.30,0.2848)
(1.40,0.1436)
(1.50,0.0176)
(1.60,0.0000)
(1.70,0.0000)
(1.80,0.0000)
(1.90,0.0000)
(2.00,0.0000)}{0.03}
\Curve{(0.10,0.0000)
(0.20,0.2104)
(0.30,0.6542)
(0.40,0.8381)
(0.50,0.9566)
(0.60,0.9810)
(0.70,0.9832)
(0.80,1.9611)
(0.90,2.7603)
(1.00,0.8505)
(1.10,0.5477)
(1.20,0.4163)
(1.30,0.2842)
(1.40,0.1434)
(1.50,0.0192)
(1.60,0.0000)
(1.70,0.0000)
(1.80,0.0000)
(1.90,0.0000)
(2.00,0.0000)}
\DashCurve{(0.10,0.0000)
(0.20,0.0836)
(0.30,0.1532)
(0.40,0.1524)
(0.50,0.1540)
(0.60,0.1449)
(0.70,0.1383)
(0.80,0.1440)
(0.90,0.1390)
(1.00,0.1097)
(1.10,0.0864)
(1.20,0.0690)
(1.30,0.0527)
(1.40,0.0338)
(1.50,0.0086)
(1.60,0.0000)
(1.70,0.0000)
(1.80,0.0000)
(1.90,0.0000)
(2.00,0.0000)}{0.013}
\SetScale{1.}\SetWidth{0.5}
\DashLine(105,250)(130,250){3}
\Text(10,250)[l]{${\mbox{pb} \over 10\,\gev}$}
\Text(20,280)[]{(b)}
\Text(135,250)[l]{$WW$}
\Line(105,235)(130,235)
\Text(135,235)[l]{All + QCD}
\Text(135,220)[l]{~~~~~+ ISR}
\DashLine(105,200)(130,200){1.3}
\Text(135,200)[l]{$g g q\bar{q}$}
\Text(0,-10)[]{0}\Text(100,-10)[]{100}
\Text(200,-10)[r]{200}
\Text(-10,50)[]{1}
\Text(-10,100)[]{2}
\Text(-10,150)[]{3}
\Text(-10,200)[]{4}
\Text(-10,250)[]{5}
\Text(100,-25)[t]{$m\p{i,j}$ (GeV)}
\end{picture}

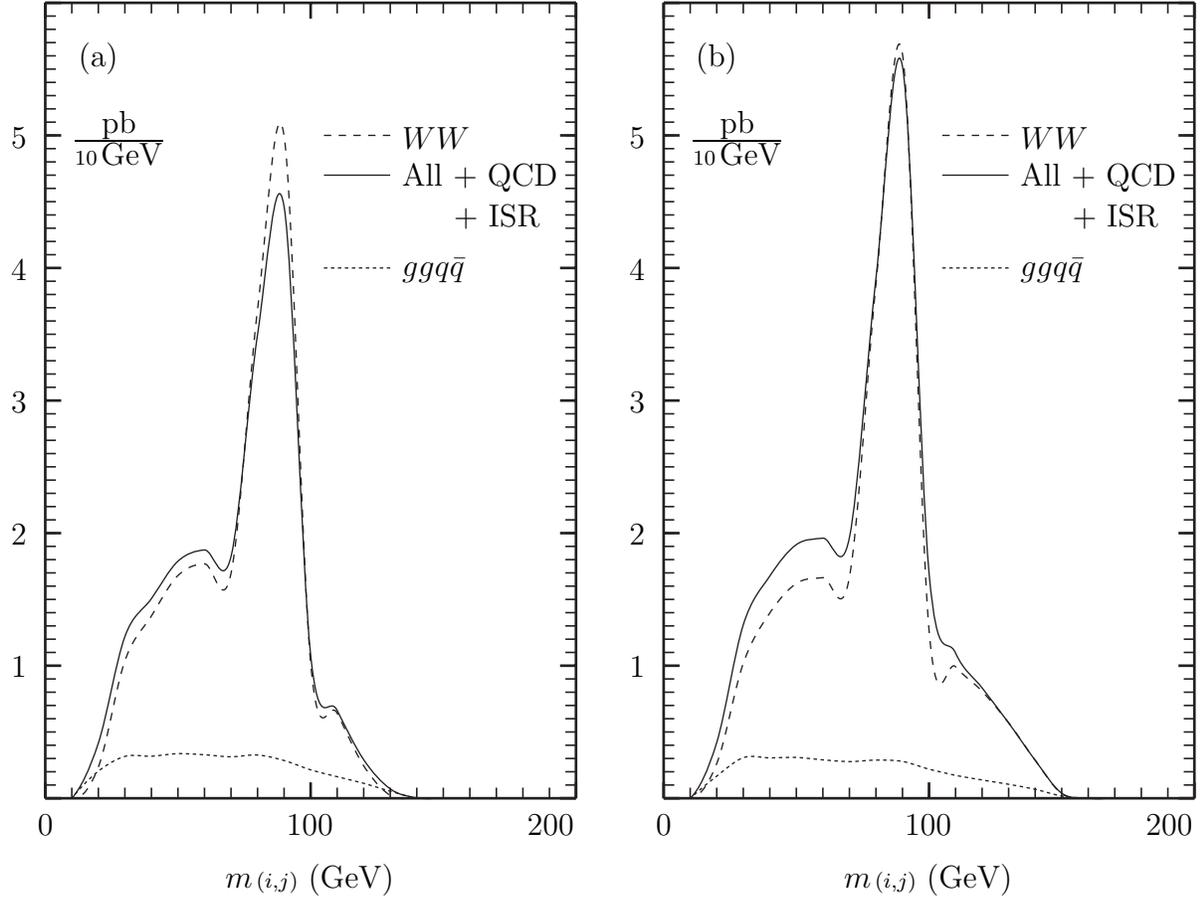
\noindent Figure 2: Distributions of the invariant masses for
$e^+ e^- \to$  4 $jets$

\noindent at $\sqrt s$ = 175 GeV (a)
 and $\sqrt s$ = 190 GeV (b). Cuts like in table 1.

\clearpage
\begin{center}
\begin{picture}(360,450)(0,8)
\LinAxis(22.5,0)(360,0)(4.5,10,5,0,1.5)
\LinAxis(22.5,500)(360,500)(4.5,10,-5,0,1.5)
\LinAxis(22.5,0)(22.5,500)(10,10,-5,0,1.5)
\LinAxis(360,0)(360,500)(10,10,5,0,1.5)
\Text(97.5,-10)[t]{{\large 79.3}}
\Text(172.5,-10)[t]{{\large 80.3}}
\Text(247.5,-10)[t]{{\large 81.3}}
\Text(322.5,-10)[t]{{\large 82.3}}
\Text(10.5,100)[r]{{\large 0.04}}
\Text(10.5,200)[r]{{\large 0.08}}
\Text(10.5,300)[r]{{\large 0.12}}
\Text(10.5,400)[r]{{\large 0.16}}
\Text(33,475)[l]{\large ${\mbox{pb} \over 100\,\mev}$}
\DashLine(97.5,100)(168,100){4}
\Text(178,100)[l]{\large $WW$}
\Line(97.5,85)(168,85)
\Text(178,85)[l]{\large All + QCD + ISR}
\DashLine(97.5,70)(168,70){2}
\Text(178,70)[l]{\large $WW$ + ISR}
\DashLine(97.5,55)(168,55){1.1}
\Text(178,55)[l]{\large $g g q \bar{q}$}
\Text(191,-30)[t]{$m\p{i,j}$ (GeV)}
\SetScale{50.}
\SetWidth{0.005}
\DashCurve{(0.45,2.5908)
(0.60,2.7788)
(0.75,2.9612)
(0.90,3.1172)
(1.05,3.2428)
(1.20,3.5180)
(1.35,3.7252)
(1.50,4.0376)
(1.65,4.3264)
(1.80,4.6520)
(1.95,4.9004)
(2.10,5.3008)
(2.25,5.7440)
(2.40,6.2612)
(2.55,6.6872)
(2.70,7.1520)
(2.85,7.6340)
(3.00,8.1520)
(3.15,8.5680)
(3.30,9.0560)
(3.45,9.3400)
(3.60,9.5400)
(3.75,9.6640)
(3.90,9.5960)
(4.05,9.4880)
(4.20,9.1720)
(4.35,8.8560)
(4.50,8.3840)
(4.65,7.8496)
(4.80,7.3168)
(4.95,6.8012)
(5.10,6.3324)
(5.25,5.8528)
(5.40,5.4468)
(5.55,5.0048)
(5.70,4.6660)
(5.85,4.3420)
(6.00,4.0120)
(6.15,3.7584)
(6.30,3.5100)
(6.45,3.2992)
(6.60,3.0840)
(6.75,2.9100)
(6.90,2.7408)
(7.05,2.5964)
(7.20,2.4464)}{0.08}
\DashCurve{(0.45,2.2320)
(0.60,2.3204)
(0.75,2.4740)
(0.90,2.5972)
(1.05,2.7600)
(1.20,2.9188)
(1.35,3.1548)
(1.50,3.3532)
(1.65,3.6304)
(1.80,3.8256)
(1.95,4.1440)
(2.10,4.4288)
(2.25,4.8412)
(2.40,5.1964)
(2.55,5.5100)
(2.70,5.9956)
(2.85,6.3588)
(3.00,6.8608)
(3.15,7.2432)
(3.30,7.5244)
(3.45,7.7544)
(3.60,7.9908)
(3.75,8.0600)
(3.90,8.0160)
(4.05,7.8264)
(4.20,7.5764)
(4.35,7.3140)
(4.50,6.9372)
(4.65,6.5140)
(4.80,6.0668)
(4.95,5.6400)
(5.10,5.2376)
(5.25,4.8448)
(5.40,4.4636)
(5.55,4.1284)
(5.70,3.8424)
(5.85,3.5792)
(6.00,3.2920)
(6.15,3.0848)
(6.30,2.8652)
(6.45,2.6860)
(6.60,2.4936)
(6.75,2.3960)
(6.90,2.2132)
(7.05,2.1260)
(7.20,2.0092)}{0.04}
\Curve{(0.45,2.4455)
(0.60,2.5405)
(0.75,2.6650)
(0.90,2.7925)
(1.05,2.9290)
(1.20,3.0975)
(1.35,3.3175)
(1.50,3.5275)
(1.65,3.8130)
(1.80,4.0140)
(1.95,4.3400)
(2.10,4.5680)
(2.25,5.0500)
(2.40,5.3800)
(2.55,5.6700)
(2.70,6.1800)
(2.85,6.5000)
(3.00,7.0100)
(3.15,7.4100)
(3.30,7.6650)
(3.45,7.9450)
(3.60,8.1550)
(3.75,8.2250)
(3.90,8.1800)
(4.05,8.0400)
(4.20,7.7050)
(4.35,7.4200)
(4.50,7.1000)
(4.65,6.7050)
(4.80,6.2400)
(4.95,5.8550)
(5.10,5.4450)
(5.25,5.0400)
(5.40,4.6045)
(5.55,4.2940)
(5.70,4.0435)
(5.85,3.7590)
(6.00,3.5045)
(6.15,3.2690)
(6.30,3.0165)
(6.45,2.8460)
(6.60,2.6630)
(6.75,2.5425)
(6.90,2.3625)
(7.05,2.2930)
(7.20,2.1860)}
\DashCurve{(0.45,0.1570)
(0.60,0.1605)
(0.75,0.1705)
(0.90,0.1655)
(1.05,0.1605)
(1.20,0.1660)
(1.35,0.1625)
(1.50,0.1560)
(1.65,0.1605)
(1.80,0.1655)
(1.95,0.1605)
(2.10,0.1625)
(2.25,0.1660)
(2.40,0.1700)
(2.55,0.1650)
(2.70,0.1650)
(2.85,0.1500)
(3.00,0.1550)
(3.15,0.1550)
(3.30,0.1600)
(3.45,0.1550)
(3.60,0.1650)
(3.75,0.1550)
(3.90,0.1550)
(4.05,0.1500)
(4.20,0.1600)
(4.35,0.1550)
(4.50,0.1650)
(4.65,0.1600)
(4.80,0.1550)
(4.95,0.1600)
(5.10,0.1550)
(5.25,0.1605)
(5.40,0.1600)
(5.55,0.1530)
(5.70,0.1525)
(5.85,0.1555)
(6.00,0.1605)
(6.15,0.1515)
(6.30,0.1510)
(6.45,0.1605)
(6.60,0.1480)
(6.75,0.1520)
(6.90,0.1490)
(7.05,0.1580)
(7.20,0.1555)}{0.022}
\SetScale{1.}\SetWidth{0.5}
\end{picture}\end{center}

\vspace{1.5cm}
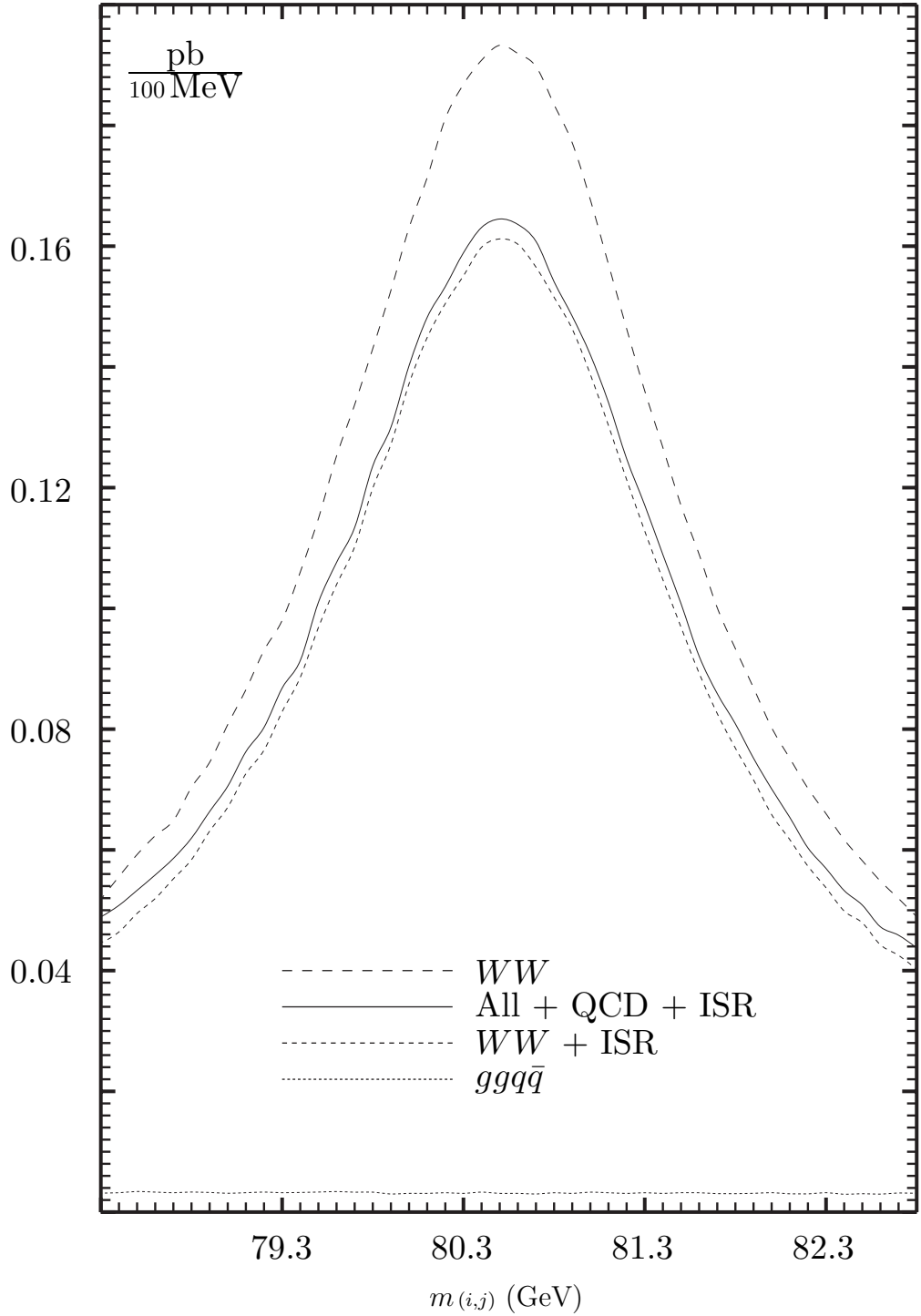
\noindent  Figure 3: Distribution of the invariant masses near the peak for

\noindent  $e^+ e^- \to$  4 $jets$ at $\sqrt s$ = 175 GeV. Cuts like in table
1.
\end{document}